
\input harvmac
%
%
\def\Title#1#2{\rightline{#1}\ifx\answ\bigans\nopagenumbers\pageno0\vskip1in
\else\pageno1\vskip.8in\fi \centerline{\titlefont #2}\vskip .5in}

scaled\magstep3
 
scaled\magstep3
%
%

%
%
\def\Fig#1{Fig.~\the\figno\xdef#1{Fig.~\the\figno}\global\advance\figno
 by1}
\def\figI{I}
%
%
\newdimen\tempszb \newdimen\tempszc \newdimen\tempszd \newdimen\tempsze
\ifx\figflag\figI
\input epsf
%
\def\epsfsize#1#2{\expandafter\epsfxsize{
 \tempszb=#1 \tempszd=#2 \tempsze=\epsfxsize
     \tempszc=\tempszb \divide\tempszc\tempszd
     \tempsze=\epsfysize \multiply\tempsze\tempszc
     \multiply\tempszc\tempszd \advance\tempszb-\tempszc
     \tempszc=\epsfysize
     \loop \advance\tempszb\tempszb \divide\tempszc 2
     \ifnum\tempszc>0
        \ifnum\tempszb<\tempszd\else
           \advance\tempszb-\tempszd \advance\tempsze\tempszc \fi
     \repeat
\ifnum\tempsze>\hsize\global\epsfxsize=\hsize\global\epsfysize=0pt\else\fi}}
\epsfverbosetrue
\fi
%
%
%
%
\def\ifigure#1#2#3#4{
\midinsert
\vbox to #4truein{\ifx\figflag\figI
\vfil\centerline{\epsfysize=#4truein\epsfbox{#3}}\fi}
\narrower\narrower\noindent{\footnotefont
{\bf #1:}  #2\par}
\endinsert
}
%
%
\font\ticp=cmcsc10
\def\sq{{\vbox {\hrule height 0.6pt\hbox{\vrule width 0.6pt\hskip 3pt
   \vbox{\vskip 6pt}\hskip 3pt \vrule width 0.6pt}\hrule height 0.6pt}}}
\def\ajou#1&#2(#3){\ \sl#1\bf#2\rm(19#3)}
\def\frac#1#2{{#1 \over #2}}
\def\eg{{\it e.g.,\ }}
\def\buildrel#1\over#2{\mathrel{\mathop{\null#2}\limits^{#1}}}
\def\hf{{1\over2}}
%
%
\lref\RSTe{J.G. Russo, L. Susskind, and L. Thorlacius, ``The endpoint of
Hawking radiation,'' Stanford preprint SU-ITP-92-17, hep-th/9206070.}
\lref\RSTc{J.G. Russo, L. Susskind, and L. Thorlacius, ``Cosmic censorship
in two-dimensional quantum gravity,''  Stanford/UT Austin preprint
SU-ITP-92-24=UTTG-19-92, hep-th/9209012.}
\lref\PresRev{J. Preskill, ``Do black holes destroy information?'' Caltech
preprint CALT-68-1819, hep-th/9209058.}
\lref\Page{D.N. Page, ``Is black hole evaporation predictable?''\ajou
Phys. Rev. Lett. & 44 (80) 301.}
\lref\Dyso{F. Dyson, Institute for Advanced Study preprint, 1976,
unpublished.}
\lref\tHoo{G. 't Hooft, ``The black hole interpretation of string
theory,''\ajou Nucl. Phys. &B335 (90) 138.}
\lref\ACN{Y. Aharonov, A. Casher, and S. Nussinov, ``The unitarity
puzzle and Planck mass stable particles,"\ajou Phys. Lett. &B191 (87)
51.}
\lref\AlSt{M. Alford and A. Strominger, ``S-wave scattering by a magnetic
black hole,''\ajou Phys. Rev. Lett. & 69 (82) 563, hep-th/9202075.}
\lref\QTDG{S.B. Giddings and A. Strominger, ``Quantum theories of dilaton
gravity,'' UCSB preprint UCSBTH-92-28, hep-th/9207034.}
\lref\liou{F. David, ``Conformal field theories coupled to 2-D gravity in
the conformal gauge,''\ajou Mod. Phys. Lett. &A3 (88) 1651\semi
J. Distler and H. Kawai, ``Conformal field theory and 2-D quantum gravity
or who's afraid of Joseph Liouville?,''\ajou Nucl. Phys. & B321 (89) 509.}
\lref\Sred{M. Srednicki, ``Is purity eternal?,'' UCSB preprint
UCSBTH-92-22, hep-th/ 9206056.}
\lref\HawkUn{S.W. Hawking, ``The unpredictability of quantum
gravity,''\ajou Comm. Math. Phys &87 (82) 395.}
\lref\Hawk{S.W. Hawking, ``Particle creation by black
holes,"\ajou Comm. Math. Phys. &43 (75) 199.}
\lref\BPS{T. Banks, M.E. Peskin, and L. Susskind, ``Difficulties for the
evolution of pure states into mixed states,''\ajou Nucl. Phys. &B244 (84)
125.}
\lref\Wald{R.M. Wald, ``Black holes, singularities and predictability,'' in
{\sl Quantum theory of gravity. Essays in honor of the 60th birthday of
Bryce S. Dewitt}, S.M. Christensen (Ed.),  Hilger (1984);
``Black holes and thermodynamics,'' U. Chicago preprint, lectures at 1991
Erice school on Black Hole Physics.}
\lref\BHMR{S.B. Giddings, ``Black holes and massive remnants,''\ajou Phys.
Rev. &D46 (92) 1347, hep-th/9203059.}
\lref\Cole{S. Coleman, ``Black holes as red herrings: Topological
fluctuations and the loss of quantum coherence,''\ajou Nucl. Phys. &B307
(88) 867.}
\lref\LIDC{S.B. Giddings and A. Strominger, ``Loss of incoherence and
determination of coupling constants in quantum gravity,''\ajou Nucl. Phys.
&B307 (88) 854.}
\lref\DXBH{S.B. Giddings and A. Strominger, ``Dynamics of Extremal Black
Holes,''\ajou Phys. Rev. &D46 (92) 627, hep-th/9202004.}
\lref\CGHS{C. Callan, S.B. Giddings, J.A. Harvey, and A. Strominger,
``Evanescent Black Holes,"\ajou Phys. Rev. &D45 (92) R1005.}
\lref\MSW{G. Mandal, A Sengupta, and S. Wadia, ``Classical solutions of
two-dimensional
string theory,'' \ajou Mod. Phys. Lett. &A6 (91) 1685.}
\lref\Witt{E. Witten, ``On string theory and black holes,''\ajou Phys. Rev.
&D44 (91) 314.}
\lref\BiDa{N.D. Birrell and P.C.W. Davies, {\sl Quantum fields in curved
space} (Cambridge U.P., 1982).}
\lref\CrFu{S. M. Christensen and S. A. Fulling, ``Trace anomalies and the
Hawking effect,''\ajou Phys. Rev. &D15 (77) 2088.}
\lref\RST{J.G. Russo, L. Susskind, and L. Thorlacius, ``Black hole
evaporation in 1+1 dimensions,'' Stanford preprint SU-ITP-92-4.}
\lref\BDDO{T. Banks, A. Dabholkar, M.R. Douglas, and M O'Loughlin, ``Are
horned particles the climax of Hawking evaporation?''\ajou Phys. Rev. &D45
(92) 3607.}
\lref\Hawktd{S.W. Hawking, ``Evaporation of two dimensional black
holes,''\ajou Phys. Rev. Lett. & 69 (92) 406,
hep-th/9203052.}
\lref\SuTh{L. Susskind and L. Thorlacius, ``Hawking radiation and
back-reaction,''\ajou Nucl. Phys & B382 (92) 123, hep-th/9203054.}
\lref\RuTs{J.G. Russo and A.A. Tseytlin, ``Scalar-tensor quantum gravity
in two dimensions,''\ajou Nucl. Phys. & B382 (92) 259.}
\lref\deAl{S.P. deAlwis, ``Quantization of a theory of 2d dilaton
gravity,''\ajou Phys. Lett. &B289 (92) 278, hep-th/9205069\semi
``Black hole physics from Liouville theory,''
Boulder preprint COLO-HEP-284, hep-th/9206020.}
\lref\BiCa{A. Bilal and C. Callan, ``Liouville models of black hole
evaporation,'' Princeton preprint PUPT-1320, hep-th/9205089.}
\lref\QBH{B. Birnir, S.B. Giddings, J.A. Harvey, and A. Strominger,
``Quantum black holes,''\ajou Phys. Rev. &D46 (92) 638,
hep-th/9203042.}
\lref\Strog{A. Strominger, ``Fadeev-Popov ghosts and 1+1 dimensional black
hole evaporation,'' UCSB preprint UCSBTH-92-18, hep-th/9205028.}
\lref\QETDBH{S.B. Giddings and W.M. Nelson, ``Quantum emission from
two-dimensional black holes,'' UCSB preprint UCSBTH-92-15,
hep-th/9204072, to appear in {\sl Phys. Rev. D.}}
\lref\GiMa{G.W. Gibbons and K. Maeda, ``Black holes and membranes in
higher-dimensional theories with dilaton fields,''\ajou Nucl. Phys. &B298
(88) 741.}
\lref\GHS{D. Garfinkle, G. Horowitz, and A. Strominger, ``Charged black holes
in string theory,''\ajou Phys. Rev. &D43 (91) 3140, erratum\ajou Phys. Rev.
& D45 (92) 3888.}
\lref\EHNS{J. Ellis, J.S. Hagelin, D.V. Nanopoulos, and M. Srednicki,
``Search for violations of quantum mechanics,''\ajou Nucl. Phys. & B241
(84) 381.}
\lref\BaOl{T. Banks and M. O'Loughlin, ``Classical and quantum production
of cornucopions at energies below $10^{18}$ GeV,'' Rutgers preprint
RU-92-14.}
\lref\BuCh{T.T. Burwick and H. Chamseddine, ``Classical and quantum
considerations of two-dimensional gravity,'' Z\"urich preprint ZU-TH-4/92.}

\Title{\vbox{\baselineskip12pt\hbox{UCSBTH-92-36}
\hbox{hep-th/9209113}
}}
{\vbox{\centerline{Toy Models for Black Hole
Evaporation{$^*$}}}}\footnote{}{*To appear in the proceedings of the
International Workshop of Theoretical Physics, 6th Session, {\sl String
Quantum Gravity and Physics at the Planck Energy Scale}, 21 -- 28 June
1992, Erice, Italy.}

\centerline{{\ticp Steven B. Giddings}\footnote{$^\dagger$}
{Email addresses:
giddings@denali.physics.ucsb.edu, steve@voodoo.bitnet.}
}

\vskip.1in
\centerline{\sl Department of Physics}
\centerline{\sl University of California}
\centerline{\sl Santa Barbara, CA 93106-9530}
\bigskip
\centerline{\bf Abstract}

These notes first present a brief summary of the puzzle of information loss to
black holes, of its proposed resolutions, and of the flaws in the proposed
resolutions.
There follows a review of recent attempts to attack this
problem, and other issues in black hole physics, using two-dimensional
dilaton gravity theories as toy models.  These toy models contain
collapsing black
holes and have for the
first time enabled an explicit semiclassical treatment of the backreaction
of the Hawking radiation on the geometry of an evaporating black hole.
However, a complete answer to the information conundrum seems to require
physics beyond the semiclassical approximation.  Preliminary attempts to
make progress in this direction, using connections to conformal field
theory, are described.

\Date{}

\vfill\eject

Since the discovery of black holes, physicists have been faced with the
possibility that they engender a {\it breakdown of predictability}. At
the classical level this breakdown arises at the singularity.
Classically we do not know how to evolve past it.  Inclusion of quantum
effects may serve as a remedy, allowing predictable evolution, by
smoothing out the singularities of general relativity. However, as
suggested by Hawking \refs{\Hawk, \HawkUn}, quantum effects also present
another sharp challenge to predictability through the mechanism of
Hawking evaporation.

To see this, consider a pure quantum state describing an infalling
matter distribution.  If this matter collapses to form a black hole, it
will subsequently emit Hawking radiation.  In Hawking's approximation
where the backreaction of the emitted radiation on the geometry is
neglected, the radiation is thermal and is described by a mixed quantum
state.  This suggests that once the black hole evaporates the initial
pure state has been converted to a final mixed state; information has
been lost, and unitarity has been violated.  Hawking proposed that this
represents a new and fundamental type of unpredictability inherent in quantum
gravity.

Beyond any prejudice that quantum gravity shouldn't violate unitarity,
there are potential problems with this scenario.  In particular, as
argued in \refs{\BPS\EHNS - \Sred}, na\"\i ve
attempts to formulate unitarity
violating dynamics typically
run afoul of essential principles such as energy
conservation.  We are therefore motivated to look for other possible
resolutions to the problem of what happens to information that falls into a
black hole.

There have been various proposals for resolving the black hole information
problem,\foot{For more comprehensive reviews see \refs{\Wald\BHMR-\PresRev}.}
but each of them appears to have flaws.  A list of these, in ascending
order of speculative content, and together with objections, is as
follows:
\item{1.} Correcting Hawking's calculation by including the backreaction
renders the final state pure; the information escapes in the Hawking
radiation.\foot{This alternative has for example been advocated in
\refs{\Page,\tHoo}.}
Objection: this would appear to imply that either all of the
information has been extracted from the infalling matter by the time it
crosses the horizon, or that information propagates acausally from
behind the horizon to outside.
\item{2.} The information is released in the last burst of radiation as
the black hole evaporates to the Planck scale and quantum-gravitational
and backreaction effects dominate. Objection: Since the initial black
hole could have been arbitrarily massive,
it must be possible for the remaining Planck scale energy to carry
off
an arbitrarily large amount of
information corresponding to all of the possible black hole initial states.
A large amount of information can be transmitted using a small amount of
energy only over a long period of time, \eg\ through emission of
many very soft
photons.  This implies the next proposal\refs{\HawkUn,\ACN,\BHMR}.
\item{3.} The black hole leaves behind a long-lived
remnant with Planck-sized mass.
This remnant must have infinitely many states to allow it to carry the
unbounded amount of information that could have been present in the
initial state.\foot{In one possible realization of this
proposal\refs{\CGHS\BDDO-\DXBH} the infinite number of states
arises from excitations of an infinite ``internal''
volume of the black hole.}
Objection: an infinite spectrum of states with Planck-size masses
wreaks havoc with loop calculations and with
production probabilities both in thermodynamics and in background fields.
In particular, such a spectrum
implies infinite production of these particles in the Hawking radiation from
arbitrarily large black holes, and likewise appears to imply
infinite production from a thermal ensemble at any given temperature.
The resulting instabilities are
disastrous.\foot{For an attempt to evade these see \refs{\BaOl}.}
\item{4.} Baby universes form and carry away the information that
falls down the black
hole, and thus unitarity, while apparently violated in our Universe,
 is restored for the system including the baby
universes\refs{\Dyso}.  Objection: in a different context, it has been argued
\refs{\Cole,\LIDC} that wormholes simply shift coupling constants and
don't lead to such apparent violations of unitarity.
\item{5.} Information emerges from the black hole via a previously
unsuspected mechanism rather than through small corrections to the Hawking
radiation.  Such a mechanism is suggested both by the
failure of other attempts to resolve the information problem and by
arguments that there should be upper bounds on information content that
arise from Planck scale physics\refs{\BHMR}.
Objection: this proposal appears to require
acausal behavior behind the horizon.

There are other variations on these basic possibilities.  The objections
are not iron-clad, and may have loopholes.  One way (and perhaps the
only way) to actually solve the information problem is to gain control
of backreaction and quantum gravity effects. This is a difficult task
in four dimensions, and it behooves us to search for simple toy models
to gain more intuition.

One such toy model\refs{\CGHS} is two-dimensional dilaton
gravity, described by the
action
\eqn\dilgrav{
S= {1\over2\pi}
\int d^2 x\  \sqrt{-g}\ \left[e^{-2\phi}\left(R+ 4(\nabla \phi)^2 +
4\lambda^2\right) - \half \sum\limits^N_{i=1} (\nabla f_i)^2\right]\ .
}
Here $\phi$ is the dilaton, $4\lambda^2$ the cosmological constant, and
the $f_i$ are minimally coupled matter fields.  Note for future
reference that since the gravitational part of the action is multiplied
by $e^{-2\phi}$, the quantity $e^\phi$ plays the role of the
gravitational coupling constant.

This toy model has several virtues.  First,
it is perturbatively
renormalizable by power counting;
the only dimensionful coupling constant is $\lambda$.
Secondly, it is completely soluble at the classical level.  Among these
solutions are black holes, and these black holes Hawking radiate
$f$-particles.  Finally, this model is the low energy effective theory
for certain types of four- and five-dimensional black holes, and this
provides a direct application to higher-dimensional physics.

Before pursuing the former points, let us recall the connection to
higher-dimensions.\foot{For a more complete explanation see \refs{\DXBH}.}
The low energy action for string theory is of the
form
\eqn\String{
S= \int d^{10}x\ \sqrt{-g}\ e^{-2\phi} \left[R + 4(\nabla\phi)^2 -{1\over2}
F^2_{\mu \nu} + \cdots + {\cal O} (\alpha^\prime)\right]
}
where $F_{\mu \nu}$ is the electromagnetic field strength and where terms
involving other fields are neglected, as are higher-dimension operators.
 This theory is known \refs{\GiMa , \GHS} to have magnetically charged
black hole solutions,
\eqn\three{
\eqalign{
ds^2 & = - \frac{1-\frac{r_+}{r}}{1-\frac{r_-}{r}}\ dt^2 +
\frac{dr^2}{\left(1-\frac{r_+}{r}\right)\left(1-\frac{r_-}{r}\right)} +
r^2 d\Omega^2_2 + ds^2_6\cr
e^{-2\phi} & = e^{-2\phi_0}\left(1 - \frac{r_-}{r}\right)\cr
F_{\mu \nu} & = Q\ \epsilon^{(2)}_{\mu \nu}\ .\cr}
}
Here $r_+$, $r_-$, and $\phi_0$ are constants, $ds^2_6$ denotes your favorite
6-dimensional string compactification, $Q$ is the magnetic charge, and
$d\Omega^2_2$, $\epsilon^{(2)}_{\mu \nu}$ are the line element and
Levi-Civita tensor on the two-sphere.

\ifigure{\Fig\pend}{The Penrose diagram for the
magnetically charged dilatonic
black hole in four dimensions.  Also shown is a constant-time slice $S$
through the geometry.}{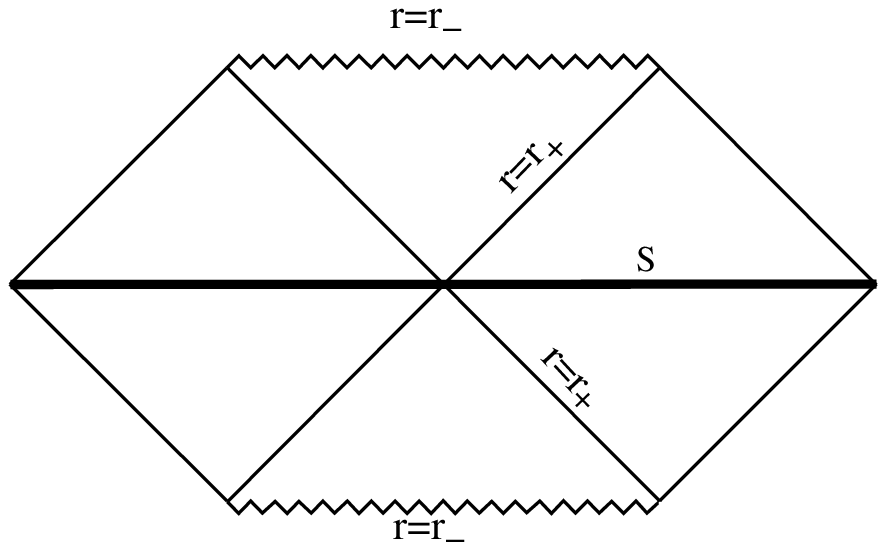}{2.75}

This solution has a causal structure identical to that of Schwarzschild
(see \pend), with a horizon at $r=r_+$
and a singularity at $r=r_-$.
However, there is a crucial
difference between their geometries.
Consider the slice $S$ shown in Fig.~1; in Schwarzschild, this spatial
slice gives the Einstein-Rosen bridge.  In the present case there is
also a throat connecting two asymptotically flat
regions (\Fig\slice), but in the
extremal limit $M\to Q$ the throat becomes infinitely long.  In this
limit an observer in the asymptotic region sees the horizon and
singularity disappear to infinity; likewise an observer fixed
near the horizon
sees the asymptotic region recede to infinity.\foot{At first sight one
might think that the extremal Reissner-Nordstrom black hole has the same
property, since its spatial geometry also has an infinite throat.  However,
the horizon does not become causally disconnected from the rest of
spacetime as a result of the rapid falloff of
$g_{00}$ along the throat.}
For the latter observer
the universe is an infinite tube terminating in a black hole; this
solution takes the form
\eqn\tdbh{
\eqalign{
ds^2 & = ds^2_{2 \rm{DBH}} (r,t) + Q^2d\Omega^2_2\cr
e^{2\phi} & = e^{2\phi_{2{\rm DBH}} (r)}\cr
F_{\mu \nu} & = Q\ \epsilon^{(2)}_{\mu \nu}\ .\cr}
}
Here $ds^2_{2{\rm DBH}}$ and $\phi_{2{\rm DBH}}$ are the metric and
dilaton for the two dimensional black hole in string theory that was
found by Mandal, Sengupta, and Wadia \refs{\MSW} and Witten
\refs{\Witt}; we will see their explicit forms shortly.
The important point is that the solution \tdbh~
is a direct product of two
two-dimensional solutions.   The second of these  is the round
two-sphere threaded by a magnetic flux.  A similar construction holds
for five-dimensional black holes.

\ifigure{\slice}{Pictured is the spatial geometry of the right
half of the slice $S$ of \pend.}{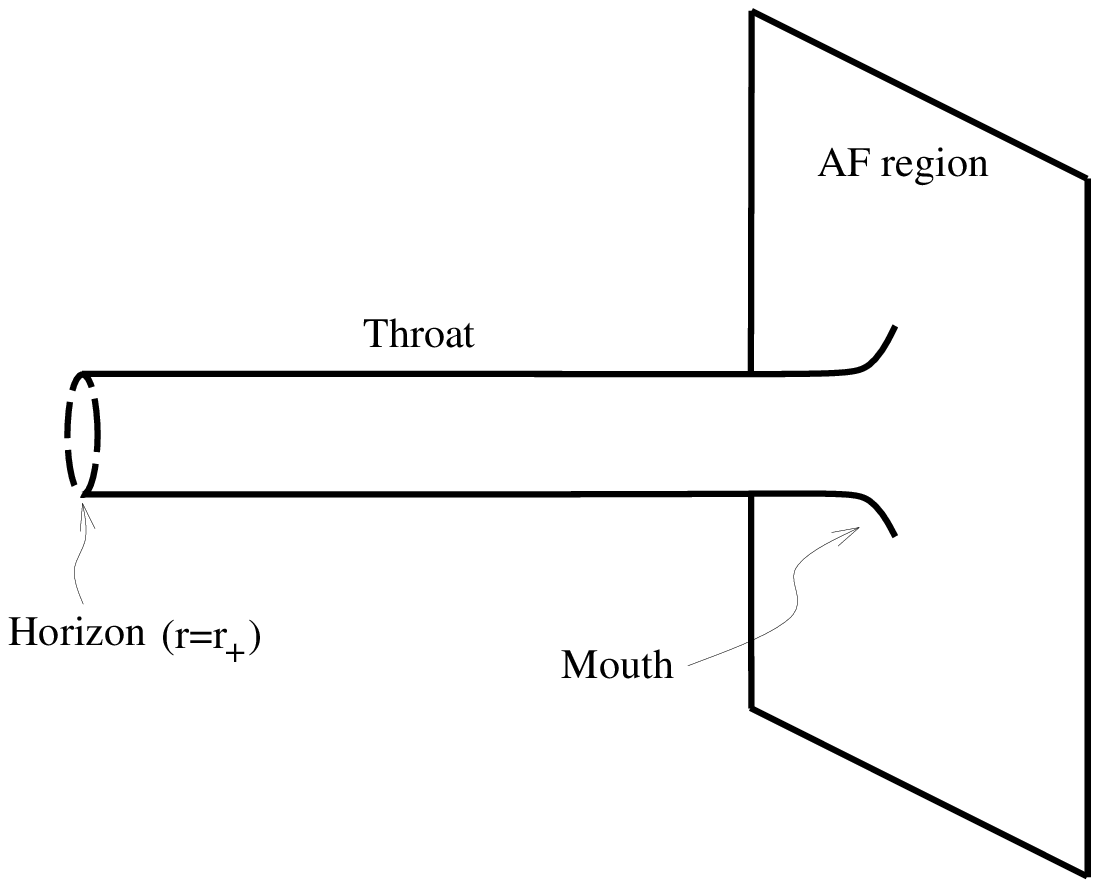}{3}

The mass of the two-dimensional black hole in \tdbh\ depends on the
asymptotic value of the dilaton, $\phi_0$.  If the latter is scaled to
$-\infty$ as the extremal limit is taken, then the mass can be arranged to
be finite.  Conversely, if the asymptotic dilaton is fixed, then the mass
is zero and the resulting two-dimensional solution is the vacuum.

With the dilaton ({\it i.e.} the coupling) fixed at infinity, and near the
extremal limit, the
solution far down the throat is closely
approximated by \tdbh\ with a small mass.
At extremality, and as seen by the asymptotic
observer, the horizon is infinitely far
down the throat  and the $r,t$ solution far down the throat is the
vacuum.
Let us now consider low-energy
scattering of particles from the
extremal black hole.  Very low
energy excitations that penetrate into the throat will not be able to
excite the angular degrees of freedom, which have a threshold $\sim
1/Q$.  On the other hand, $s$-wave excitations may penetrate the
throat and raise the mass of the black hole above extremality.  This
corresponds to raising the mass of the two-dimensional black hole
above zero.  The excess mass will later be emitted in Hawking
radiation, corresponding to evaporation of the two-dimensional
solution back to zero mass.  This provides a direct relationship
between low-energy scattering by near-extremal dilatonic black holes
and formation and evaporation of two-dimensional black holes.  The
low-energy effective theory describing such excitations is obtained
by dropping the angular dependence in \String, and one obtains the
action \dilgrav.\foot{To describe excitations that can actually
penetrate the throat, one must add matter terms to the action
\String.  This is further described in \refs{\DXBH,\AlSt}.}

To investigate the two-dimensional problem we return to the action
\dilgrav.  Although the general solution is easily found \refs{\CGHS}, we focus
on
some special cases; units are chosen so that $\lambda=1$.  First are the
vacuum solutions, with $f_i\equiv 0$:
\eqn\five{
\eqalign{
ds^2& = -\frac{dx^+ dx^-}{M-x^+ x^-}\cr
e^{-2\phi} & = M-x^+ x^-\cr}
}
where $x^\pm = x^0 \pm x^1$ are light-cone coordinates and $M$ is an
arbitrary parameter.
The change of coordinates $x^+ = e^{\sigma^+}\ ,\ x^-=-e^{-\sigma^-}$ (with
$\sigma^\pm = \tau\pm\sigma$)
gives
\eqn\six{
\eqalign{
ds^2 & = -\frac{d\sigma^+ d\sigma^-}{1+Me^{\sigma^--\sigma^+}}\cr
\phi &= -\half \ell n\left(M + e^{\sigma^+ - \sigma^-}\right)\ .\cr}
}
For $M=0$ the metric is flat, and this solution is identified as the
ground state of the theory.  Note that the dilaton is then
\eqn\lduf{ \phi = -\sigma;
}
the $M=0$ solution is therefore called the {\it linear dilaton vacuum}.
For $M>0$ one recovers the two-dimensional black hole of \refs{\MSW,
\Witt}.  The causal structure of this black hole is identical to that of
Schwarzschild; its Penrose diagram is given by \Fig\tdpen.
Solutions with $M<0$ have naked singularities.

\ifigure{\tdpen}{The Penrose diagram for the two-dimensional eternal black
hole.}{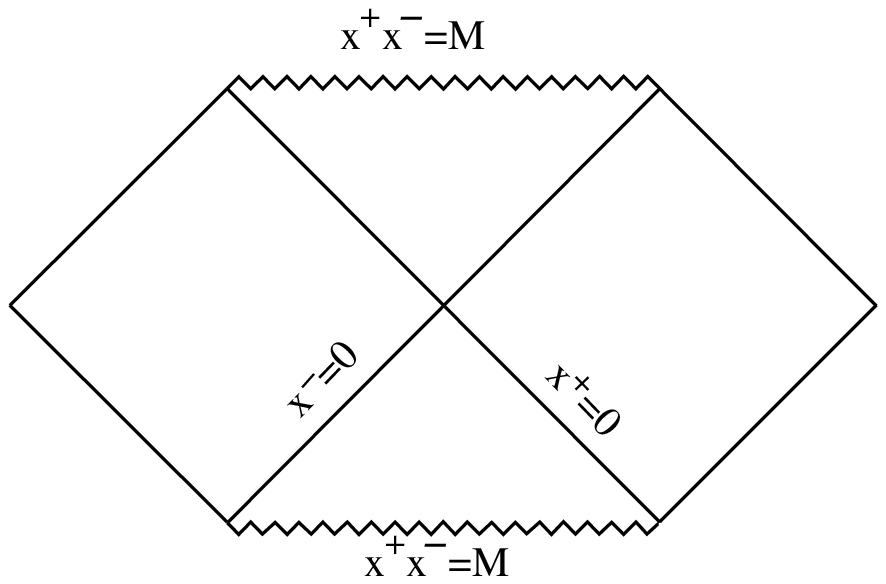}{2.75}

Black hole formation occurs when one allows matter to fall into the
linear dilaton vacuum. For example, a general
left-moving lump of
classical matter is given by
\eqn\eight{
f=F(x^+)\ ,
}
for an arbitrary function $F(x^+)$, and has stress tensor
\eqn\strt{T_{++}=\hf \left(\partial_+ F\right)^2\ .}
An $F(x^+)$ that vanishes
outside $x^+_f > x^+ > x^+_i$, or equivalently
$\sigma^+_f>\sigma^+>\sigma^+_i$, corresponding to a lump of finite width,
 yields a Penrose diagram as in \Fig\infall: the matter
``collapses'' to form a black hole.
The metric for $\sigma^+ <
\sigma^+_i$ is
\eqn\fmet{
ds^2 =- d\sigma^+ d\sigma^-
}
and for $\sigma^+> \sigma^+_f$ is
\eqn\bhcoll{
ds^2 = -\frac{d\sigma^+ d\sigma^-}{1+M\,e^{\sigma^- - \sigma^+} - \Delta
\, e^{\sigma^-}}\  .
}
Here the constants $M$ and $\Delta$ are moments of the matter distribution,
\eqn\moments{M= \int_{\sigma_i^+}^{\sigma_f^+} d\sigma^+ T_{++}(\sigma^+)\
,\ \Delta= \int_{\sigma_i^+}^{\sigma_f^+} d\sigma^+ e^{-\sigma^+}
T_{++}(\sigma^+)\ .}
The change of coordinates
\eqn\eleven{\xi^+=\sigma^+\ ,\
\xi^- = - \ell n\left[e^{-\sigma^-} - \Delta\right]
}
returns \bhcoll~ to the asymptotically flat form,
\eqn\twelve{
ds^2 = -\frac{d\xi^+ d\xi^-}{1 + M\,e^{\xi^- - \xi^+}}\ .
}

\ifigure{\infall}{The Penrose diagram for an ``infalling'' lump of
classical matter.  Also indicated are the ``in'' and ``out'' regions for
right movers.}{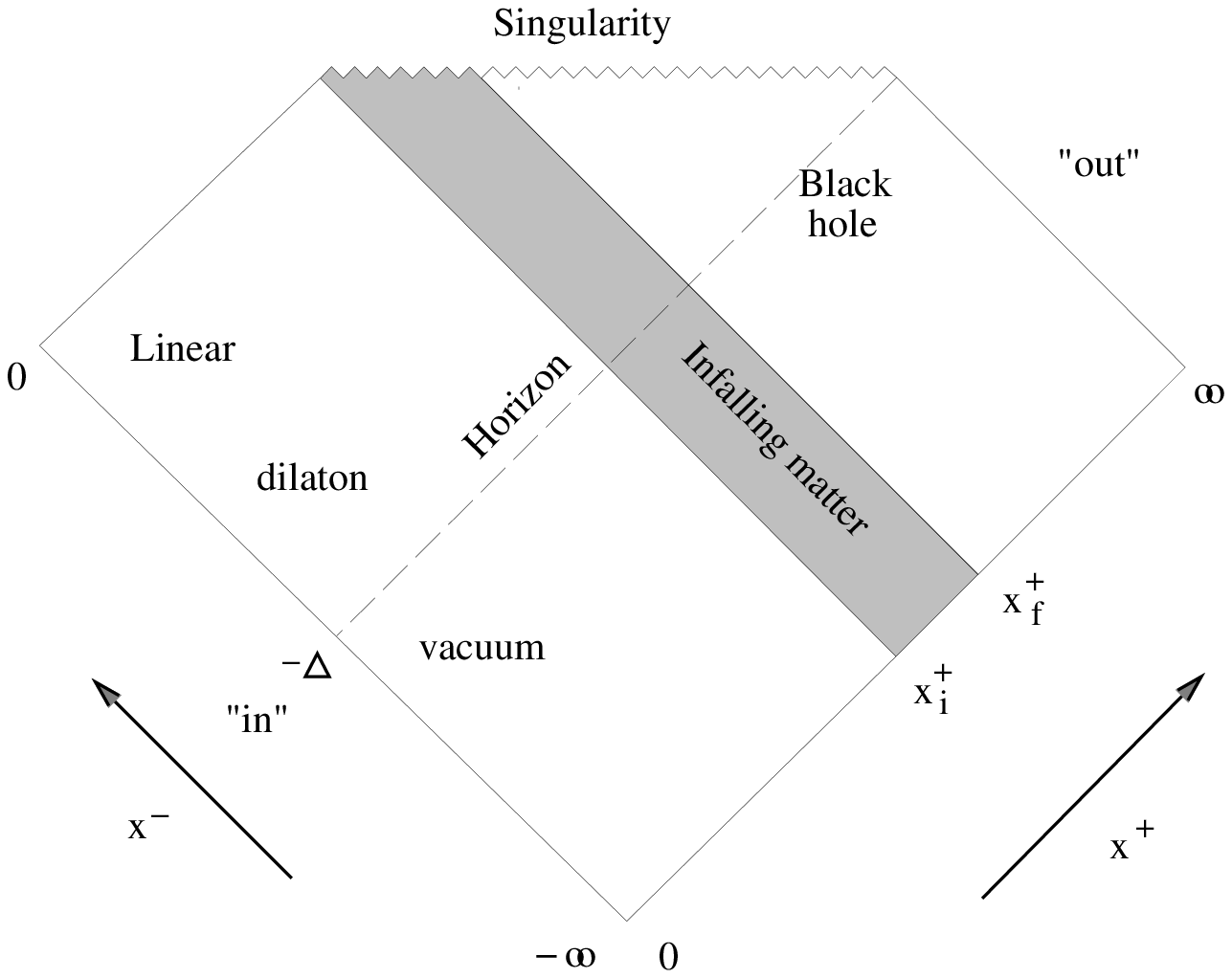}{3.5}

Hawking radiation arises from evolution of positive frequency incoming
states to both positive and negative frequency outgoing states.\foot{For
more details, see \eg \refs{\BiDa,\QETDBH}.}
In a fixed background the left and right moving $f$-quanta are decoupled,
and in considering the Hawking radiation we focus on the right-movers.
The positive frequency modes in the
respective regions are
\eqn\thirteen{
\eqalign{
u_\omega & = \frac{1}{\sqrt{2\omega}}\ e^{-i\omega\sigma^-}\qquad
 {\rm (in)}\cr
v_\omega & = \frac{1}{\sqrt{2\omega}}\ e^{-i\omega \xi^-}\qquad {\rm (out)}\cr}
}
and they are related by the Fourier transform,
\eqn\fourteen{
v_\omega \left(\xi^-(\sigma^-)\right) = \int\limits^\infty_0 d\omega^\prime
\left[\alpha_{\omega\omega^\prime} u_{\omega^\prime} (\sigma^-) +
\beta_{\omega\omega^\prime}
u^*_{\omega^\prime} (\sigma^-)\right]\ .
}
The Fourier coefficients $\alpha_{\omega\omega^\prime},
\beta_{\omega\omega^\prime}$ give
the Bogoliubov transformation, and determine the spectrum of Hawking
radiation.  For example, it is easily shown that the expectation value
of the out number operator in the in vacuum is given by
\eqn\fifteen{
{}_{\rm in}\left\langle 0\left|N^{\rm out}_\omega
\right|0\right\rangle_{\rm in}
= \int\nolimits^\infty_0 d\omega^\prime\left|
\beta_{\omega\omega^\prime} \right|^2\ ;
}
nonvanishing $\beta_{\omega\omega^\prime}$ therefore implies Hawking radiation.

In the present case the Bogoliubov coefficients can be evaluated exactly
\refs{\QETDBH} (contrast four-dimensional black holes!); for example,
\eqn\sixteen{
\beta_{\omega\omega^\prime} = \frac{1}{2\pi}
\ \sqrt{\frac{\omega^\prime}{\omega}}
\ \Delta^{i\omega} B\left(-i\omega-i\omega^\prime, 1+i\omega\right)
\ .
}
{}From this one can derive \refs{\QETDBH} the outgoing stress tensor,
\eqn\stresst{
\left\langle T^f_{--}\right\rangle = \frac{1}{48} \left[1 -
\frac{1}{\left(1+\Delta\,e^{\xi^-}\right)^2}\right]
}
which, after an initial transitory period, describes a thermal flux of
Hawking radiation.

Next we wish to investigate the backreaction of the Hawking radiation on
the geometry.  It is most easily discussed by considering the vacuum
functional integral,
\eqn\eighteen{
\int {\cal D}g {\cal D}\phi\ e^{iS[\phi, g]} \int {\cal D}_g f
\ e^{-{i\over4\pi}\int d^2x\,\sqrt{-g}\
\sum\limits^N_{i=1} (\nabla f_i)^2}\ .
}
Here we have split the action into the gravitational part and the  matter
part.  The subscript on the matter measure indicates that this
functional measure is induced from the functional metric
\eqn\fmetric{
\left(\delta f_1, \delta f_2\right)_g = \int d^2x\ \sqrt{-g}\ \delta f_1
\ \delta f_2
}
in the usual fashion.  The matter functional integral has been
extensively studied in the string literature and elsewhere, and gives
\eqn\eighteena{
\int {\cal D}_g f\ e^{-{i\over4\pi}\int d^2x\,\sqrt{-g}\
\sum\limits^N_{i=1}(\nabla
f_i)^2} = e^{iNS_{\rm PL}}
}
where
\eqn\polliv{
S_{PL} = - \frac{1}{96\pi} \int d^2x_1 \sqrt{-g} \int d^2x_2 \sqrt{-g}
\ R(x_1) \sq^{-1} (x_1, x_2) R(x_2)
}
is the Polyakov-Liouville action. (Here $\sq^{-1}$ denotes the Green
function for the d'Alembertian.) Although this action is nonlocal, it
appears local in conformal gauge, $ds^2 = -e^{2\rho} dx^+
dx^-$:
\eqn\pollivuc{
S_{PL} = {1\over24\pi} \int d^2x\ (\nabla\rho)^2\ .
}

Gravitational dynamics, including quantum effects of the matter,
therefore follow from the functional integral
\eqn\fint{
\int {\cal D}g {\cal D}\phi\ e^{iS[g, \phi] + iNS_{PL}[g]}
\ .
}
One way to see the gravitational effects of the matter is to compute the
quantum stress tensor,
\eqn\twentythree{N{2\pi\over\sqrt{-g}}
\frac{\delta S_{PL}}{\delta g^{\mu \nu}} =
\left\langle T^f_{\mu
\nu}\right\rangle\ .
}
One finds either directly from \polliv, or equivalently from the
well-known trace anomaly,
\eqn\trace{
\left\langle T^f_{+-}\right\rangle = -{N\over96}R = -\frac{N}{12}
\ \partial_+\partial_-\rho\ .
}
The other components of the stress tensor can also be found by varying
\polliv, although it is simpler to integrate the conservation law
$\nabla^\mu\langle T^f_{\mu \nu}\rangle  = 0$ using \trace.  This
gives
\eqn\twentyfive{
\left\langle T^f_{--}\right\rangle = -\frac{N}{12}
\left[(\partial_-\rho)^2 - \partial_-^2\rho + t_-(x^-)\right]
}
where $t_-(x^-) $ is an integration function that must be fixed by
boundary conditions.  (Equivalently it arises from the ambiguity in
defining the Green function in \polliv).

Computation of $\langle T^f_{--}\rangle$ from this expression and
\bhcoll~ yields agreement with \stresst, confirming the relationship
\refs{\CrFu} between the conformal anomaly and Hawking radiation.  We
conclude that $S_{PL}$ incorporates both the Hawking radiation, and the
effects of its backreaction on the geometry.

What, then,  are these effects?  To begin, let's determine when the
backreaction due to the $f$-fields has a substantial effect on the
geometry.  This can be estimated by asking when the Hawking radiation
(uncorrected by backreaction effects) has carried away a substantial
fraction of the mass of the black hole,
\eqn\twentysix{
\int d\xi^- \left\langle T^f_{--} \right\rangle \sim M\ .
}
It is straightforward to see that for large $M$
this occurs at the retarded time
$x^-_{\rm evap}\simeq-\Delta$, near the horizon, and where
the dilaton
at the trailing edge of the incoming
matter distribution satisfies
\eqn\twentyseven{
e^{2\phi} \sim {1\over M}\ .
}
Therefore if the mass of the black hole is taken to be large (and also $x_i^+
\Delta \gg 1$), the evaporation process takes place at weak coupling.  This
helps us in separating the backreaction from quantum-gravitational effects.

%

Indeed, the resulting weak-coupling can be used to justify a
semi-classical analysis of the functional
integral \fint, via the classical equations arising from the action
\eqn\semic{
S_{SC} = {1\over2\pi}
\int d^2x\ e^{-2\phi}\left[-2\,\sq\rho + 4(\nabla\phi)^2 +
4\lambda^2\right] + \frac{N}{24\pi} \int d^2x (\nabla\rho)^2
}
where we use the conformal-gauge result $R=-2\,\sq\rho$.
In order to do this we must take the number of matter fields $N$ to
be large so that the second term
dominates the other quantum corrections to the dynamics, and can be treated
on the same footing as the first term.

Because the correction term in \semic~ is quadratic in $\rho$, and
$\rho=0$ for the linear dilaton vacuum, it remains a solution to the
semiclassical theory.  However, the theory is no longer exactly soluble
and, in fact, other solutions are difficult to find.  We have,
nonetheless, gained considerable insight into the structure of the
solutions from numerical and general arguments
\refs{\RST\BDDO\Hawktd\SuTh-\QBH}.

As an example, from the resulting time-independent equations one can
numerically find static solutions\refs{\Hawktd\SuTh-\QBH}
that correspond to a black hole in
equilibrium with an influx of radiation that precisely balances the outward
Hawking flux.

In the dynamic case of a black hole formed from collapsing matter,
one can argue that an apparent horizon, determined by the
condition $(\nabla\phi)^2=0$, forms and recedes
as the black hole
evaporates. A surprise is that behind this apparent horizon is a
singularity of the semiclassical equations\refs{\RST,\BDDO}.
The reason this is
surprising is that it is distinct in behavior from the original
classical singularity; in particular, it occurs where
\eqn\thirty{
e^{2\phi} = e^{2\phi_{cr}} = \frac{12}{N}\ .
}
It is therefore (for large $N$)
not at large $e^\phi$ as was the classical singularity.
Mathematically this singularity arises as a result of the vanishing of
an eigenvalue of the kinetic operator in \semic; this signals a
breakdown of the semiclassical approximation.  This breakdown means that
solutions of the equations following from \semic~ should really not be
trusted for values of the dilaton $\phi \geq \phi_{cr} -\epsilon$,
for
some small ($N$-dependent) $\epsilon$.

\ifigure{\Fig\krusk}{A Kruskal diagram for the evaporating two-dimensional
black
hole.  $Q$ denotes the line along which $\phi=\phi_{cr} -\epsilon$; beyond
this line a full quantum description of the collapse is presumably needed
to make predictions.}{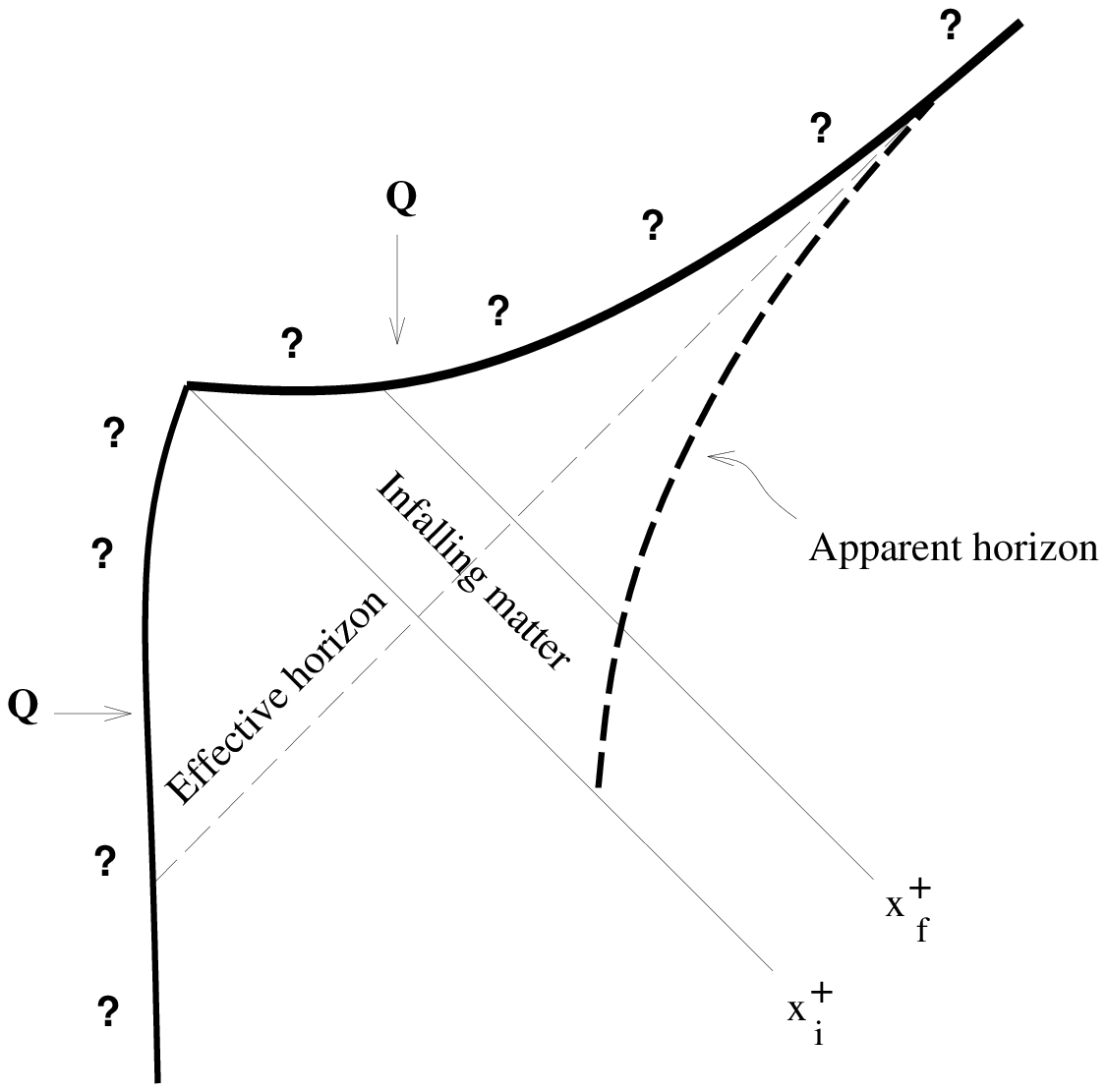}{3.7}

\ifigure{\Fig\qpen}{A Penrose diagram for the evaporating two-dimensional
black hole.  Prediction of physics in the causal future of the line $Q$
requires a quantum treatment of dilaton gravity.  In particular, one cannot
at present determine the final fate of the black hole.}{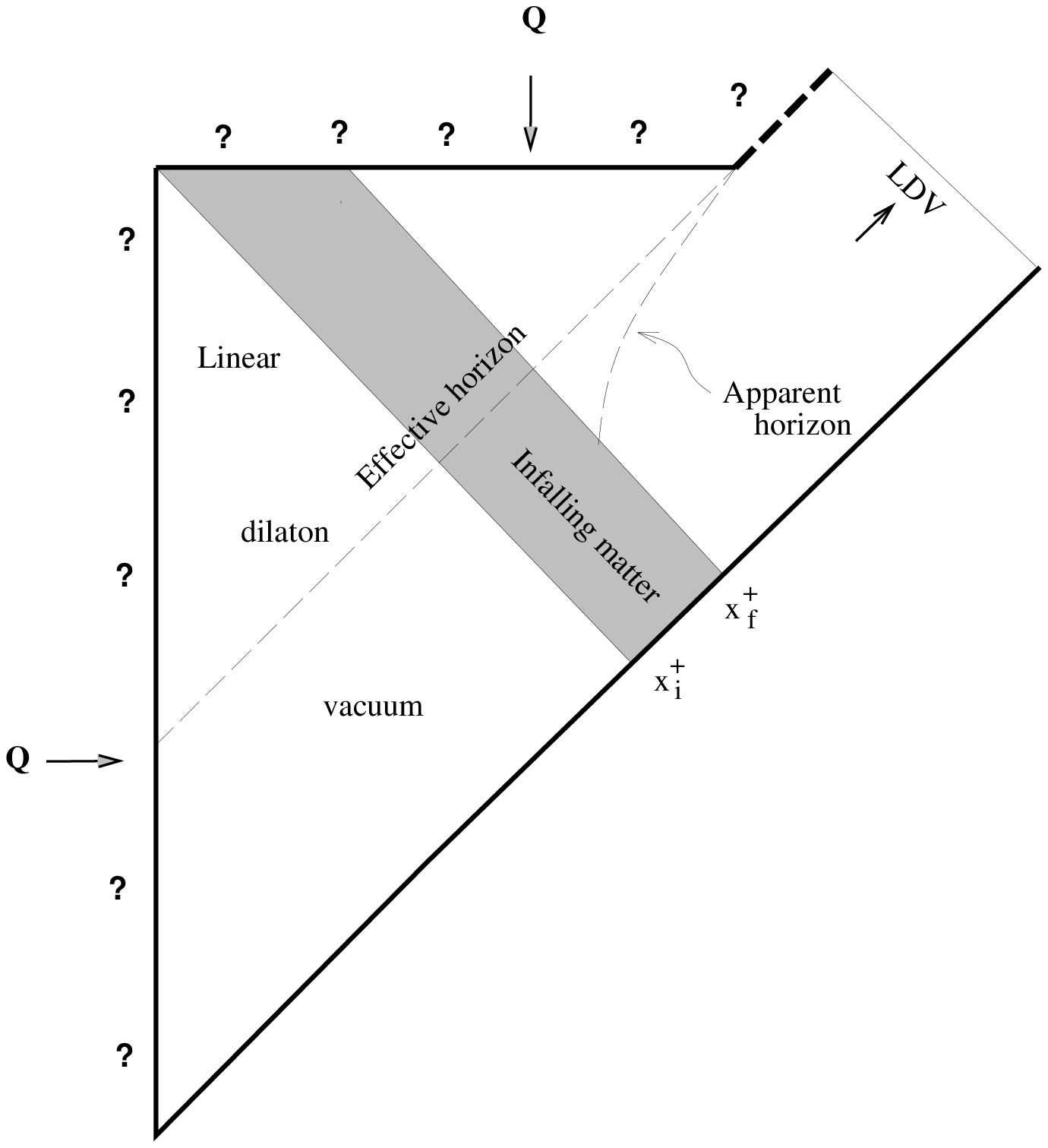}{4}

The resulting geometry is pictured in \krusk, \qpen.  Shown are both the
apparent horizon, and the {\it effective horizon}.  The latter will in
general be defined as the boundary of the region
from which future-directed causal curves can escape to null infinity
without encountering Planck-scale physics; it is thus the boundary of the
region where we can make predictions using an effective field
theory valid below the Planck scale.
In the present context ``Planck scale''
physics is identified with physics beyond the validity of the semiclassical
approximation.
Eventually the line with
$\phi=\phi_{cr}-\epsilon$ crosses the horizons. We cannot determine the
physical behavior in the future light cone (or ``shadow'') of this
point  since it depends on the presently ill-understood physics at
$\phi \geq \phi_{cr}-\epsilon$.  For $\phi < \phi_{cr} -\epsilon$, the
solution asymptotes back to the linear dilaton vacuum.

This semiclassical picture, although incomplete, may nonetheless give
some clues about the information problem.  In particular, working
order-by-order in $1/N$, it is probable that one can
construct an argument\refs{\QETDBH}
that information does not come out in the
Hawking radiation before one reaches the shadow of $\phi = \phi_{cr}
-\epsilon$.  This is analogous to stating that the information
doesn't come out of four-dimensional black holes until they reach the
Planck scale. It may therefore, in the present context, rule out the most
conventional proposed resolution, resolution 1), of the black hole
information conundrum.

On the other hand, semiclassical predictability has failed and we still
can't say what {\it does} happen to information.  To proceed we must go beyond
the large-$N$ approximation and investigate the full quantum theory.

Quantum dilaton gravity is most easily treated by gauge-fixing the
metric to the form
\eqn\indecomp{
g_{\mu \nu} = e^{2\rho} \hat g_{\mu \nu}
}
where $\hat g_{\mu \nu}$ is a fixed background metric.   Although we have
argued that dilaton gravity is renormalizable, the fields $\rho$ and
$\phi$ are dimensionless so in fact an infinite number of counterterms
can occur.  The general parity invariant action is of the form
\eqn\sact{
S= -\frac{1}{2\pi} \int d^2x\ \sqrt{-\hat g} \ \left(G_{MN} (X^P) \hat{\nabla
X^M\, \cdot\, \nabla X^N} + \half\Phi (X^P) \hat R + T (X^P) \right)
}
where $X^P = (\rho, \phi)$.  Here $G_{MN} (X^P)$, $\Phi (X^P)$, and
$T(X^P)$ are arbitrary functions, and we use sigma-model notation.

In quantizing dilaton gravity there are several physical restrictions on
the general action \sact.  The most obvious of these are:
\itemitem{1)} The theory should depend on $\hat g, \rho$ only through
the combination in \indecomp; that is, the theory should be invariant
under the background transformation
\eqn\thirtythree{
\eqalign{
\hat g_{\mu \nu} & \to e^{2\omega} \hat g_{\mu \nu}\cr
\rho &\to \rho - \omega \ .\cr}
}
This condition is on-shell equivalent to invariance under conformal
rescaling of the background metric $\hat g$, which implies that
the
sigma-model $\beta$-functions must vanish.  If we momentarily reinstate
Planck's constant, and work to leading order in $\hbar$, this implies
\eqn\bfcn{\eqalign{& \nabla_M\Phi \nabla^M T -4T -{\hbar\over2}\sq T =0\cr
&\nabla_M\nabla_N \Phi + {\hbar\over2} R_{MN}=0\cr
&\left(\nabla \Phi\right)^2 -{\hbar\over2} \sq\Phi+ {(N-24)\hbar\over3}=0\
.}}
Off-shell these must be supplemented by the condition that the tangent
vector $V^M$ to the $\rho$ direction satisfy $V_M = \nabla_M\Phi/2$.
\itemitem{2)} The theory must agree with dilaton gravity at weak
coupling:
\eqn\zloop{
\eqalign{
G_{MN} & \lower11pt\hbox{$ \buildrel\textstyle\to
\over {\scriptstyle e^\phi\to 0}$} G^{c\ell}_{MN}=
\left(\matrix{-4e^{-2\phi} & 2e^{-2\phi}\cr
        2e^{-2\phi} & 0\cr}\right)\cr
\Phi & \to \Phi^{c\ell} = -2e^{-2\phi} \cr
T & \to T^{c\ell} = -4 \lambda^2 e^{-2\phi}\cr}
}

 Furthermore, outside the large-$N$ approximation we must
worry about the role of the ghosts.  If the measures for the ghost and
$(\rho, \phi)$ functional integrals are defined using the metric $g$, as
in \fmetric, then this results in the replacement $N\to N-24$ in \semic.
This implies the nonsensical result that for $N<24$, black holes
accrete mass by Hawking radiating negative energy in ghosts!  The
problem is easily resolved by instead using the metric $e^{2\phi} g$ to
regulate the functional measures \refs{\Strog, \QTDG}. There follows a
third condition:
\itemitem{3)} At weak coupling, subleading counterterms of the form
\eqn\oneloop{
\eqalign{
G_{MN} & \lower11pt\hbox{$\buildrel\textstyle\to
\over {\scriptstyle e^\phi\to 0}$} G^{c\ell}_{MN} +\hbar
\left(\matrix{2  &-2\cr
       -2 & \frac{24-N}{12}\cr}\right)+\cdots\cr
\Phi & \to \Phi^{c\ell} + \hbar\frac{24-N}{6} \,\rho -4\hbar\phi + \cdots
\ \cr}
}
should appear in $G_{MN}$, $\Phi$.

 Finally there may be other physical constraints; one such
restriction is
\itemitem{4)} The theory should have a sensible ground state.

Writing down the full $\beta$-function equations \bfcn, let alone
solving them, is no small task.  One promising approach has been
advocated in refs.~\refs{\RuTs\BuCh\deAl-\BiCa, \QTDG}. As is easily seen,
the leading order metric given in \oneloop~ is flat, and therefore
trivially obeys the leading-order $\beta$-function equations.
Furthermore, if $T=0$ this theory is an exact CFT, that is, an exact
solution of the $\beta$-function equations.

One can then identify the tachyon as the operator of conformal dimension
(1,1) that agrees with $T^{c\ell}$ to leading order in $e^\phi$.  The
theory with $T\not= 0$ is obtained by perturbing the exact flat theory
with this operator.  This is similar to steps used to define Liouville
theory \refs{\liou}, and should yield an exact solution to the
$\beta$-function equations.

Although the resulting theory satisfies criteria 1--3, it does not
satisfy criterion 4.  It can be shown that there are regular solutions
with mass unbounded from below \refs{\QTDG}. Hawking radiation in these
theories does not shut off \refs{\BiCa, \deAl}, and  black holes appear
to radiate to infinite negative mass.  The necessary modifications for a
stable theory are not obvious;  one attempt  to stabilize such models is
by applying suitable boundary conditions at the line where
$\phi=\phi_{cr}$\refs{\RSTe,\RSTc}.

Despite these facts, the general approach of attempting to identify
exact conformal field theories that represent evaporating black holes is
worthy of pursuit; perhaps other more realistic examples can be
constructed.  One is still, however, left with the feeling that
uniqueness is lacking.  Consideration of supersymmetric theories
may provide sufficient uniqueness and solve the problem of negative
mass.  A different tack is to view the problem of the non-uniqueness of
quantum dilaton gravity as similar to that of four-dimensional gravity.
In the latter case, we expect string theory to provide an escape from
non-renormalizability.  Perhaps two-dimensional dilaton gravity is best
treated as the low-energy limit of string theory as well.

To conclude, we have succeeded in qualitatively understanding
two-dimensional black hole formation and evaporation until quantum
effects become strong; this is analogous to understanding
four-dimensional black holes up to the Planck regime.  Furthermore, we
may likely rule out the most conservative proposed resolution to the
black hole information problem.  This is potentially very interesting.
However, a solution to the information conundrum is still beyond the
horizon.  We haven't seen quantum restoration of predictability, and
probably won't until we understand quantization of the {\it family} of
theories of quantum dilaton gravity.
Although this is a challenge, it is a good toy problem to develop
techniques for higher-dimensions: quantization of dilaton gravity should
be an excellent warmup for understanding quantum gravity in four
dimensions.

\bigskip\bigskip\centerline{{\bf Acknowledgements}}\nobreak
I wish to thank my collaborators B. Birnir,
C. Callan, J. Harvey, W. Nelson, and A. Strominger for a stimulating
collaboration.  I have also benefitted from discussions with S. Hawking,
J. Preskill,
L. Susskind, and L. Thorlacius.  This work was supported in part by
DOE grant DE-FG03-91ER40168 and NSF
PYI grant PHY-9157463.

\listrefs

\end